\documentclass[aps,twocolumn,superscriptaddress,amssymb,amsmath]{revtex4-2}
\usepackage{graphicx}
\usepackage{hyperref}
\usepackage{appendix}
\usepackage{csquotes}
\usepackage{color}

\newcommand{\e}{\mathrm{e}}
\renewcommand{\d}{\mathrm{d}}

\newcommand{\etal}{\textit{et~al.}}
\newcommand{\av}[1]{\left\langle#1\right\rangle}

\begin{document}

        \title{Hierarchical Balance Theory: Emergence of Instability in Follower Layer Below Critical Temperatures}
        \author{Amir Kargaran}
	\email{amir.kargaran@ipm.ir}
	\affiliation{School of Biological Sciences, Institute for Research in Fundamental Sciences (IPM), 19395-5746, Tehran, Iran}
        \author{Houman Jafari}
	\affiliation{Electrical Engineer, Sharif University of Technology, P.O. Box 11365-9161, Tehran, Iran}
	\author{G.Reza Jafari}
	\email{g\_jafari@sbu.ac.ir}
	\affiliation{Physics Department, Shahid Beheshti University 1983969411, Tehran, Iran}
	\date{\today}
	
    \begin{abstract}
    Hierarchy significantly shapes interactions in social structures by organizing individuals or groups based on status, power, or privilege. This study investigates how hierarchy affects structural balance as temperature variations, which measure an individual's average irrationality in society. To address this question, we develop a two-layer balance model, the \enquote{leader layer}, which maintains structural balance exclusively through intra-layer interactions. Conversely, the \enquote{follower layer} maintains structural equilibrium through both inter- and intra-layer interactions. The Hamiltonian of the leading layer is independent, while the follower layer depends on its parameters as well as those of the leading layer. Analytical results from the mean-field approximation and exact Monte Carlo simulations show that instability arises in the equilibrium states of the follower layer when the temperature is below the critical threshold ($T<T_c$), which is different from the structural Heider equilibrium. Furthermore, our findings indicate that the critical temperature is elevated in the follower layer.
    \end{abstract}
    
    \maketitle
    \section{\label{sec:level1}Introduction}
    Heider first introduced the balance theory to model the social tension between three individuals \cite{heider1}\cite{heider2}. The combination of this model with graph theory \cite{cartwright}, makes this model powerful, and its impact is so vast that we can find the usage of this model in social networks \cite{szell,altafini,samin,gallo1,gallo2}, international relations \cite{hart,galam,bramson}, biology \cite{Yen-Sheng,majid2,abbas,nastaran} and ecology \cite{saiz}. In the analytical approach, researchers usually consider a single layer, and all individuals have class equity, and the concept of \enquote{social inequality} is ignored in the modeling. To explore this influence, we define a more realistic balance theory model with a hierarchy containing the leader and follower layers.

    Hierarchy is a fundamental concept in both the natural and social sciences. It refers to systems in which elements are ranked according to their status or power \cite{mageeh}. This organizational structure is typical in human societies as well as biological contexts, such as animal dominance hierarchies \cite{drews}. Scientists consider this concept to model complex systems. For instance, Clauset \etal \cite{newmanhirarchy} presents a method for inferring hierarchical structures in networks. They show that this organization explains key network characteristics like degree distributions, clustering, and path lengths. Additionally, they demonstrate that understanding hierarchical structure can predict missing connections in partially known networks with high accuracy. Similarly, Magneta \cite{mantegna} examined financial markets and uncovered a hierarchical organization among stocks, which helps understand market behaviors and anomalies. Ravasz \etal \cite{barabási1,barabási2} demonstrated that modularity, which is often observed in metabolic and other biological networks, can coexist with a scale-free topology in complex networks. An interesting area of research in hierarchical networks is the information propagation and mechanisms that create hierarchical networks \cite{paluch}. Siudem \etal analyzed diffusion dynamics on hierarchical systems of weakly-coupled networks, revealing that modifying the network structure can alter diffusion constants \cite{siudem}. Czaplicka \etal found that information limitations in growing multi-level systems can significantly restrain the emergence of new hierarchy levels \cite{czaplicka1} and how noise enhances information transfer in hierarchical networks \cite{czaplicka2}. This finding indicates that hierarchical modularity consists of multiple modules nested within one another, creating increasingly complex layers. The interplay between hierarchical structures and traditional social models, such as structural balance, offers an intriguing area for study.
    
    Structural balance views society as a network of individuals (nodes) connected by social ties (links) that can be positive (friendly) or negative (hostile). Triads, or interactions among three individuals, are categorized into four types: $[+ + +]$, $[+ - -]$, $[- - -]$, and $[+ - +]$. The first two types are balanced, while the latter two are imbalanced. Balanced triads create less social tension, described by the phrases "a friend of my friend is my friend" and "the enemy of my enemy is my friend." Two stable states characterize overall balance: "heaven," where all triads are positive ($[+++]$), and "bipolar," where individuals in one subgroup have friendly ties ($[+++]$) while opposing those in another subgroup ($[+ - -]$) \cite{cartwright}. Researchers are investigating the structural balance of real-world social data, driven by the recent wave in available real social data. For instance, Facchetti \etal \cite{facchetti} identified a measure of balance using the spin glass model, which revealed that real-world networks are generally quite balanced. However, in contrast, Estrada and Benzi \cite{estrada} proposed a walk-based balance measure indicating that online networks are, surprisingly, poorly balanced. Leskovec et al. \cite{leskovec} conducted evaluations of theories related to signed networks, including structural balance, using online datasets. They present an alternative model for explaining the contradictory outputs from data using the structural balance model in evolving directed networks. Some scientists choose another approach and redefine the concept of balance as a spectrum instead of a binary parameter, as noted by \cite{estrada,kirkely}. While these studies are interesting and show potential, they also reveal meaningful inconsistencies, suggesting that we may need to utilize more realistic models.
    
    In recent decades, researchers, especially physicists, have increasingly studied social networks as complex systems through statistical mechanics methods. This research primarily falls into two categories: non-equilibrium \cite{antal1,antal2,malarz1,malarz2,shojaei,sonubi,caram} and equilibrium \cite{belaza1,belaza2,fereshteh,amir1,amir2,farideh2,hakim1,hakim2} statistical mechanics approaches. The non-equilibrium approach summarizes balance theory in rate equations, and the stationary state of this equation is discussed with Ku\l akowski \etal \cite{kulakowski} being the first to introduce such equations, examining how link value distributions and equilibrium time vary with network size. Antal \etal \cite{antal1,antal2} expanded this by analyzing rate equations considered for all types of triads in balance theory and calculating the stationary states for all triad types concerning the free parameters of their model. Marvel \etal \cite{marvel2} highlighted the initial number of positive links as crucial in a network, particularly when starting from random conditions, is crucial in determining whether the network will eventually reach bipolar or heaven states. Recently, scientists have argued that dyadic homophylic interactions are more fundamental than structural balance. They consider a vector of attributes for individuals and look at structural balance as an emergent phenomenon. For instance, G{\' o}rski \cite{gorski2} work reveals that achieving a fully balanced state is challenging unless tremendous attributes are considered. 
    
    In equilibrium approaches, scientists consider energy functions or the Hamiltonian to investigate the system's equilibrium state. Marvel \etal \cite{marvel1} examined an energy function related to social balance and employed statistical mechanics to analyze the properties of local minima in the energy landscape, known as jammed states. Belaza \etal \cite{belaza1} developed a more complex Hamiltonian, considering the role of inactive links \cite{belaza2}. Rabbani \etal \cite{fereshteh} introduced the thermal balance theory. They investigated the phase transition in balance theory related to temperature, which reflects individuals' average irrationality. Their findings indicate that there is a discrete phase transition from order to disorder, along with the existence of a hysteresis loop. Building on the ideas presented by G{\'o} rski \cite{gorski2}, researchers are exploring a shift from the traditional equilibrium approach to structural balance. They suggest integrating personal attributes into their Hamiltonians and considering structural balance a natural phenomenon. Korbel \etal \cite{korbel} and Pham \etal \cite{pham2} have defined a clustered base Hamiltonian derived from spin glass theory. They successfully predicted the distribution of average group sizes and types of triads using data from a multiplayer online game.

    Along with extending the structural balance model, scientists consider a bi-layer or multi-layer model in which the interaction between and within layers is different. They investigate this extension in both the none-equilibrium and equilibrium approaches. As our knowledge extends, G{\'o}rski \etal \cite{gorski1} was the first to use this idea to generalize the balance theory rate equation introduced in Ku\l akowski \etal \cite{kulakowski} to the bi-layer networks where links are coupled to the intra-layer triads (Heider balance interaction) and inter-layer links (Ising model interaction) and find that Heider balance is rarely possible due to these inter-layer connections. Others explore bi-layer Hider balance in equilibrium approaches like \cite{mohandas1} and define a Hamiltonian mixture of Heider balance and Ising model Hamiltonians on fully connected networks. They find that critical temperatures of the bi-layer model are higher than those of single-layer setups. Furthermore, another work \cite{mohandas2} found how critical single and layer networks scaled with connection probability in Erd\H{o}s-R\'{e}nyi graphs. Until now, all works have had different kinds of interaction within and between different layers. To date, our research on bi-layer Heider balance has examined two types of interactions: intra-layer and inter-layer. Typically, the intra-layer interactions follow a triplet Heider balance model, while the inter-layer interactions are symmetric dyadic. However, we believe hierarchy affects factors beyond symmetrical inter-layer social ties, highlighting an unexplored gap in social modeling.

     To address this gap, we present a two-layer balance theory that incorporates asymmetric triplet interactions.The leader layer consists exclusively of Heider balance interactions within itself. In contrast, the follower layer includes both intra-layer and inter-layer Heider balance interactions. This means a link in the follower layer interacts with both triads within its own layer and triangles in the leader layer. In other words, the leader layer influences the balance within the follower layer, while the leader layer remains independent of the follower layer. This relationship resembles the hierarchical structures commonly found in social networks \cite{hirarchy1,hirarchy2}.

    The structure of this paper starts with section \ref{model}, where we review the mathematics of two approaches from statistical mechanics: non-equilibrium and equilibrium. Next, we define our model along with its rate equations, employing a straightforward decoupling method. In section \ref{analysis}, we describe the transition from a non-equilibrium to an equilibrium approach, explaining the equivalent Hamiltonian and applying a mean-field approximation. The mathematical framework of our analytic approach is the statistical physics, especially \textit{exponential random graph } \cite{robins,snijders,newman1,newman2,book}, that we use to find mean value quantities as a function of temperature, which is the uncertainty that happens randomly in societies. Then, we compare our mean-field approximation with simulation results and examine the instability found in the follower layer outcomes and the discrepancies between the critical temperatures of leaders and followers.

    \section{Model}\label{model}
    The following section reviews single-layer balance theory equilibrium and non-equilibrium approaches. The last subsection presents our two-layer balance theory's definition.
    \subsection{Non-equilibrium approach to single-layer balance theory}\label{first_sub_model}
    K. Ku\l akowski \etal \cite{kulakowski} based on a single link dynamic equation introduced the none-equilibrium statistical mechanics approach to balance theory. This equation explains how a single link, a social relationship between individuals, evolves concerning all the triangles to which it belongs. The number of triangles associated with each link in an all-to-all connected network is $n-2$, where $n$ represents the total number of nodes in the network. The social relationship evolved the dynamic equation as 
    \begin{equation}\label{kula-dynamic}
    \frac{d\sigma_{ij}}{dt}= \sum_{k=1}^{n-2} \sigma_{ik}\sigma_{jk},
    \end{equation}
    where $\sigma_{ik}$ is a link between node $i$ and $k$, the value can be a positive and negative real number. The sign of link, positive/negative, is related to the type of relationship, which can be friendly/hostile between two agents, and the weight corresponds to the power of the relationship. For example, the relationship is extremely friendly/enmity if the weight is a significant positive/negative real number. This dynamic pushes the system into what in non-equilibrium literature is called the stationary states, where the system's meaningful average does not change much with time. The dynamic described in Eq.~\ref{kula-dynamic} has two different stationary states: \enquote{heaven} or \enquote{utopia} and a \enquote{bipolar} state, which fundamentally differ from each other. In the heaven state, all links are positive, and all agents have friendly relationships. In contrast, in the bipolar state, the system is divided into two sub-groups: within each group, all links are friendly, and links between the groups exhibit hostility or negativity.

    \subsection{Equilibrium Approach to Single-Layer Balance Theory}\label{second_sub_model}
    Scientists explore the single-layer balance theory using equilibrium statistical mechanics. This approach defines the energy function or, more generally, the Hamiltonian to examine the equilibrium states \cite{marvel1} and \cite{fereshteh}. The energy of the single-layer balance theory is the sum of all network balance and imbalance triangles, with a negative sign encouraging the system to reach balanced states. The triangles in the balance theory definition is the multiplication of three links that belong to a triangle. The Hamiltonian for single-layer balance theory is defined as:
    \begin{equation}\label{fereshteh-ham}
    \mathcal{H}(G) =-\sum_{i<j<k}\sigma_{ij}\,\sigma_{jk}\,\sigma_{ki}.
    \end{equation}
    
    In this context, like before, the notation \(\sigma_{ij}\) represents the social tie between two agents ($i$ and $j$). Unlike the previous subsection, the modeling here treats the weights of social relationships as equal. Instead, only the signs of these relationships can vary, meaning that  $\sigma_{ij}\in\{-1,1\}$. In the single-layer balance theory literature, many works discuss the equilibrium states, like before heaven, bipolar, and jammed states. Scientists tried to define more realistic modeling to explain the existence of unbalanced triangles in real social data, so they defined and added agent opinions, social temperature, or different network topologies to their extended models.

    The Social temperature was added to the equilibrium approach to this model for the first time by Rabbani \etal \cite{fereshteh}, and they wrote a mean-field approximation for this model to predict the critical temperature of the phase transition. The phase transition in single-layer balance theory is a first-order transition, indicating that in an all-to-all connected network, the system abruptly shifts from the ordered phase (heaven) to the disordered phase (random) as the temperature increases. They confirmed this transition through the Monte-Carlo simulation, demonstrating that both the mean-field approximation and the simulations are in good agreement.
        
    The mean-field approximation involves approximating the mean field experienced by each link from all neighboring interacting parts, which, in balance theory, are triangles. This approximation is accomplished by summing all terms related to a specific link, such as $\sigma_{ij}$, and using mean-field theory to approximate the remaining terms. In line with this approach, we can express the single-layered balance Hamiltonian from Eq. \ref{fereshteh-ham} as $\mathcal{H} = \mathcal{H}' + \mathcal{H}_{ij}$, where \(\mathcal{H}_{ij}\) includes all terms in the Hamiltonian that involve $\sigma_{ij}$, and $\mathcal{H} '$ comprises the remaining terms. Thus, we can write:
    \begin{equation}\label{mean-field1}
    \mathcal{H}_{ij}=- \sigma_{ij}{\sum_{k\neq{i,j}}\sigma_{jk}\sigma_{ki}},
    \end{equation}
    The mean value $\langle \sigma_{ij}\rangle$ can be calculated as
    \begin{equation}\label{mean-field2}
        \langle{\sigma_{ij}}\rangle =\frac{1}{\mathcal{Z}}\sum_{\{\sigma'\neq \sigma_{ij}\}}e^{-\beta\mathcal{H}'}\sum_{\sigma_{ij}={\left\{\pm{1}\right\}}}\sigma_{ij}e^{-\beta\mathcal{H}_{ij}},\\
    \end{equation}
    where $\mathcal{Z} = \sum_{_{\left\{ G \right\}}}\exp{(-\beta\mathcal{H}(G))}$ is the partition function and $\beta$ is inverse of temperature. By this method, Rabbani et al \cite{fereshteh} find the mean-field approximation of average links $p\equiv\langle\sigma_{ij}\rangle$ and two-stars $q\equiv\langle\sigma_{ij}\sigma_{jk}\rangle$ for a fully connected network with a number of nodes (size) of $n$. A two-star configuration features a central node that connects to two other nodes, creating a V-shape. Finally, the authors derive a self-consistency equation for two stars as $q \equiv f(q; \beta, n)$ using these mean quantities, which is crucial as it encapsulates all equilibrium information. By solving this equation, we can find the stable and unstable solutions versus the model's free parameters, like temperature and size. The graphical solution of this self-consistency equation is discussed in this paper, and under a critical temperature ($T<T_c$), there is one stable solution where the mean of two-stars is one. The solutions suddenly become three when the temperature crosses this critical value ($T>T_c$). This solution is two stable and one unstable solution. The stable solutions correspond to the ordered and random states of the system. The sudden shift in the number of solutions and the presence of the \enquote{hysteresis loop} indicates that this phase transition is discrete, recognized in the literature as a first-order phase transition.

    The equilibrium and non-equilibrium approaches discussed in earlier sections have advantages and disadvantages. One major drawback of the equilibrium approach is its inability to predict bipolar situations due to violating the homogeneity assumption inherent in the mean-field theory. Additionally, in the non-equilibrium approach, there is no straightforward analytical method for solving the rate equations; thus, simulations are the only viable option.
    \begin{figure}[t]
        \centering
        \includegraphics[width=1\linewidth]{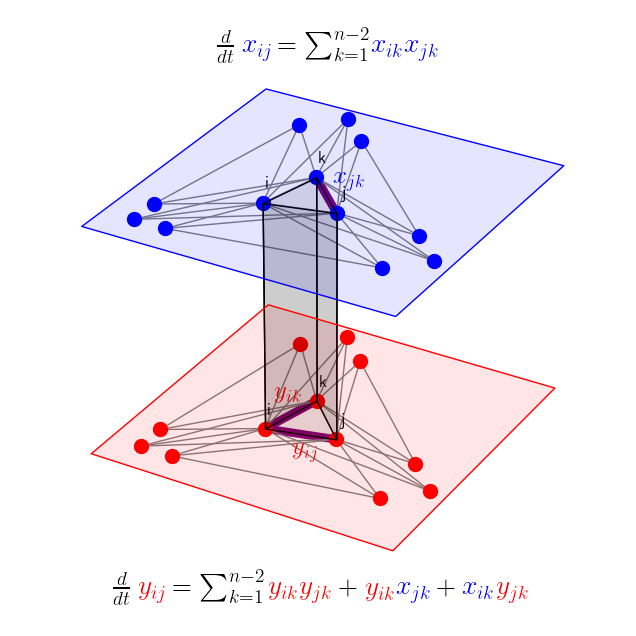}
        \caption{The schematic representation of our two-layer balance theory. The blue layer denotes the leader layer, while the red layer represents the follower layer. The leader (follower) layer links are $x$ ($y$). All connections in the leader layer interact solely with intra-layer triads. In contrast, the follower layer connections engage with both inter-layer and intra-layer triads.}
        \label{fig:main1}
    \end{figure}
    \subsection{Two-Layered Balance Dynamics Equations} \label{third_sub_model}
    In this subsection, we aim to define our model, which consists of a two-layer balance theory involving specific interactions. The leader layer functions independently and does not interact with the follower layer. In this layer, all links evolve with triangles that belong to this layer. Conversely, the follower layer is influenced by the leader layer, and each connection in this layer must consider the triangles within its own layer and those from the leader layer. In other words, the dynamic of the follower layer is coupled to the leader layer and defines a hierarchy Fig.~\ref{fig:main1}. We called links in the leader layer by $x$ and links in the follower layer named by  $y$, and the dynamic equations for both layers is
    \begin{equation}\label{rateEq}
       \left\{
       \begin{aligned}
            \frac{\d x_{ij}}{dt}&= \sum_{k=1}^{n-2} x_{ik}x_{jk}\\
        \frac{\d y_{ij}}{dt}&= \sum_{k=1}^{n-2} (y_{ik}y_{jk} + y_{ik}x_{jk} + x_{ik}y_{jk} ). \\
        \end{aligned}
        \right.
    \end{equation}
     As before, the $x_{ij}$ is the name of a link between two agents $i$ and $j$. We consider signed networks, where the link values for both layers are $x_{ij}, y_{ij}\in \{-1,1\}$. The dynamic equation for the leader layer is comparable to the work of K. Ku\l akowski \etal \cite{kulakowski}, which indicates that all social ties exhibit inter-layer interactions. In contrast, the follower layer contains inter-layer interactions and intra-layer interactions (Table.~\ref{tab:table}). For instance, let us consider an arbitrary link in the follower layer, denoted as $y_{ij}$. This link should consider all two-stars within its layer $\sum_{k} y_{ik}y_{jk}$ as well as two types of mixed two-stars involving links from both layers ($\sum_{k}  y_{ik}x_{jk}$ and $\sum_{k} x_{ik}y_{jk}$). In other words, the balance in the follower layer is influenced by the leader layer, and recognizing this balance hierarchy is a crucial key to generating more realistic models in social science.

    To begin solving the coupled dynamic equations for a two-layer balance, we can add the right-hand side and left-hand side of Eq.~ \ref{rateEq} together and find
    \begin{equation}
        \frac{\d x_{ij}}{\d t}+\frac{\d y_{ij}}{\d t}=\sum_{k} (x_{ik}x_{jk} + y_{ik}y_{jk} + y_{ik}x_{jk} + x_{ik}y_{jk}).
    \end{equation}

    \begin{table}[t]
        \centering
        \includegraphics[width=1\linewidth]{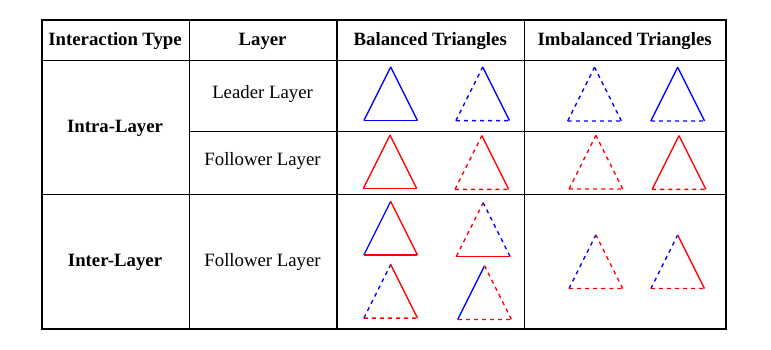}
        \caption{This table shows our model's inter- and intra-layer triplet interactions. The friendly/hostile interactions in the leader layer are in blue solid/dashed lines, and the follower layer is in red solid/dashed lines. Both leader and follower layers have inter-layer interaction, which resembles the traditional balance theory. However, the follower layer has additional interaction, which is defined as intra-layer interaction in our model. Each link in this layer interacts with mixed two-stars, resulting in more triangles.}
        \label{tab:table}
    \end{table}
    
    We can decouple this set of equations by making a straightforward change of variables. If we define $z_{ij}=(x_{ij}+y_{ij})/2$, and by considering
    \begin{equation}\label{total-two-star}
            \begin{aligned}
                 z_{ik}z_{jk}&=(x_{ik}+y_{ik})(x_{jk} + y_{jk})/4\\
                 &= (x_{ik}x_{jk} + x_{ik}y_{jk} +y_{ik}x_{jk} + y_{ik}y_{jk})/4
            \end{aligned}
	\end{equation}
    we can write the decoupled form of the two-layer balance theory model as $\d z_{ij}/dt = 2\sum_{k}z_{ik}z_{jk}$. The rate equation presented above is the decoupled format of Eq.~\ref{rateEq}, leading to a single equation that describes the total evolution of the system. This change in variables significantly impacts the balance theory for the total layers ($z$), where the values of the links can be categorized as friendly, hostile, or neutral, denoted as $z_{ij} \in \{-1, 0, 1\}$. In the literature, edges with neutral values are referred to as inactive links, and researchers have focused on models based on these links \cite{belaza2}. By considering this equation beside the leader layer ($x$) rate equation, we have two decouple rate equations as
    \begin{equation}\label{decoupleRateEq}
       \left\{
       \begin{aligned}
            \frac{\d x_{ij}}{\d t} &= \sum_{k=1}^{n-2} x_{ik}x_{jk},\\      \frac{\d z_{ij}}{\d t} &= 2\sum_{k=1}^{n-2}z_{ik}z_{jk}.
        \end{aligned}
        \right.
    \end{equation}

    \section{Analysis}\label{analysis}
    \subsection{Mean-Field Approximation for Two-Layered Balance Theory}\label{first_sub_analysis}
    
      This section will discuss the mean-field approximation applied to the two-layer balance theory. The rate equations for the two-layer balance model, as described in Eq.~\ref{decoupleRateEq}, are independent of one another. Our primary focus is on analyzing the stationary or equilibrium states of the system. In the context of equilibrium statistical mechanics, equilibrium states are defined as states in which the key quantities of the system remain constant over time. To examine these equilibrium states, we will replace the dynamic equations governing the system with their equilibrium equivalent: the Hamiltonian. As mentioned in subsection (\ref{second_sub_model}), the Hamiltonian for single-layer balance theory is governed by Eq. \ref{fereshteh-ham}. Consequently, for the two-layer system described by the dynamic equations in Eq. \ref{rateEq}, there are two Hamiltonians: one for the leader layer, denoted as $x$, and one for the total layers, denoted as $z$. We can write
    \begin{equation}\label{HamEqus}
       \left\{
       \begin{aligned}
            \mathcal{H}_x(G) &=-\sum_{i<j<k}x_{ij}\,x_{jk}\,x_{ki},\\
        \mathcal{H}_z(G') &=-2\sum_{i<j<k}z_{ij}\,z_{jk}\,z_{ki},
        \end{aligned}
        \right.
    \end{equation}
    where $\mathcal{H}_x(G)$ refers to the Hamiltonian for an leader layer defined as $x$, while $\mathcal{H}_z(G')$ represents the Hamiltonian for the total layers $z$. The phase spaces of the total layers and the leader layer differ significantly due to the presence of inactive links ($z_{ij}=0$) in the total layers. As a result, the phase spaces of each Hamiltonian are distinct. We use $G$ and $G'$ to represent the phase spaces that each Hamiltonian governs.
    
    We have two distinct Hamiltonians: one for the leader layer ($x$) and another for the total layers ($z$). We can determine equilibrium solutions for the leader and total layers using the mean-field approximation described in the previous subsection (\ref{second_sub_model}). These solutions will also provide insights into the behavior of the follower layer ($y$) in terms of mean quantities, which is the primary focus of this work. This behavior is particularly intriguing because it allows us to explore an open system (the follower layer) analytically or investigate hierarchical balance theory.
    
    The mean-field approximation for the leader layer resembles the work of Rabbani \etal \cite{fereshteh}. To analyze the total layers, we start with the average links of the total layer denoted as $\langle z_{ij} \rangle$. Using the definition of $z_{ij}$, we can express the average as $\langle z_{ij}\rangle=\left(\langle x_{ij}\rangle+\langle y_{ij}\rangle\right)/2$. Here, $ \langle x_{ij} \rangle$ and $\langle y_{ij} \rangle$ represent the averages of links for the leader and follower layers, respectively. As an essential part of the mean-field approximation within thermalized balance theory is the need to calculate the average of the two-stars in the total layer as well (Eq.~\ref{total-two-star}), we have
    \begin{equation}\label{means}
        \langle z_{ij}z_{jk}\rangle=(\langle x_{ik}x_{jk}\rangle + \langle x_{ik}y_{jk}\rangle + \langle y_{ik}x_{jk}\rangle + \langle y_{ik}y_{jk}\rangle)/4.
    \end{equation}
    
    To simplify our analysis, we will denote the average number of links with parameter $p$. The mean number of edges in the leader layer is referred to as $p_x \equiv\langle x_{ij} \rangle$, in the follower layer as $p_y \equiv \langle y_{ij} \rangle$, and the overall mean across both layers as $p_z \equiv \langle z_{ij}\rangle=(p_x + p_y)/2$. Furthermore, we will define the mean number of two-stars across all layers of the total system as follows:
    \begin{equation}\label{definitions}
        \begin{aligned}
            q_x \equiv\langle x_{ik}x_{jk}\rangle,&\quad q_{y} \equiv\langle y_{ik}y_{jk}\rangle,\quad q_{z}\equiv\langle z_{ik}z_{jk}\rangle.\\
            q_{xy}&\equiv\langle y_{ik}x_{jk}\rangle = \langle x_{ik}y_{jk}\rangle.\\
        \end{aligned}
    \end{equation}
    
    Based on the provided definition, we can express the average number of links and two-stars for the total layers as follows   
    \begin{equation}\label{definitions}
        p_z = (p_x + p_y)/2,\quad q_z=(q_x+ 2 q_{xy} +q_y)/4,
    \end{equation}
    and by utilizing the definitions above and applying techniques from statistical mechanics, as described in the previous subsection (\ref{second_sub_model}), we can derive the mean-field approximation to compute the average quantities of the Hamiltonians for both the leader layer and the total layers (Eq.~\ref{HamEqus}) as a function of temperature ($\beta=1/T$) and size ($n$). This process leads us to two self-consistent equations:
     \begin{equation}\label{selfcons}
               \left\{
               \begin{aligned}
                    q_{x} &=f(q_{x};\beta,n),\\
                    q_{z} &=g(q_{z};\beta,n).
                \end{aligned}
                \right.
    \end{equation}
                
    The derivation of the equations mentioned above can be found in Appendix \ref{appendix:self-con-fun}. By solving the self-consistency equations, we can determine the mean of the two-stars in the follower layer ($y$), according to the Eq.~\ref{definitions} and by employing the mean-field approximation where $q_{xy} \approx p_x p_y$, we can calculate the mean links and two-stars of the follower layer by
    \begin{equation}
       \left\{
       \begin{aligned}
            p_y &= 2p_z - p_x,\\
            q_y &= (4q_z-q_x)-2p_x(2p_z-p_x).
        \end{aligned}
        \right.
    \end{equation}

    Another essential concept is the stability of the solution of a self-consistence equation. We aim to establish mathematical conditions that differentiate between stable and unstable fixed points. To achieve this, we consider a point very close to a fixed point represented as $q_x^*+\delta q_x$ where $q_x^*$ is the fixed point for the first self-consistence equation in Eq.~\ref{selfcons} and $\delta q_x$ is a slight deviation with the limit $\delta q_x\to 0$. By substituting this point into \( f(q_x; \beta, n) \), we can apply the Taylor expansion of the equation. We have 
    \begin{equation}
		\begin{aligned}
                q_x^* + \delta q_x&= f(q_x^* +\delta q;\beta, n)\\
			&\approx f(q_x^*;\beta, n) + f'(q_x^*;\beta, n)\,\delta q_x,
        \end{aligned}
    \end{equation}
    where in above, we neglect the higher order of $\delta q_x$ because the deviation is very small. If the $|f' (q_x^*;\beta, n)|<1$, we can conclude that the fixed point is stable. Conversely, the fixed point is unstable if this condition is not met. The reason for that can be seen easily from a recursive perspective. The right-hand side of the above equation becomes smaller/larger by each iteration if the condition is satisfied/unsatisfied. In our model, the leader's and total layers' stability is calculated directly from their self-consistency equations. The follower layer's stability condition is satisfied if both fixed points in the $x$ and $z$ layers are stable; otherwise, it is considered unstable.
    
    \subsection{Simulations}\label{second_sub_analysis}
    We conduct our simulations on an all-to-all network consisting of $n$ nodes. Our simulation focuses on finding mean quantities, such as the average two-stars, and the energy of configurations that minimize the system's energy (Appendix.~\ref{appendix:energy-follower-layer}). To obtain the mean quantities of these configurations, we must minimize the system's energy, given an initial condition at a specific temperature. The Monte Carlo simulation is one of the most effective methods. We employed the thermalized version of this method in our work. The central component of the thermalized Monte Carlo simulation is the Metropolis algorithm. This algorithm flips a given randomly selected link or, in other words, changes its sign. This process creates a new configuration network. If the energy of the new configuration decreases, the flipped is accepted. If the energy increases, the flipped may still be accepted with a probability determined by the Boltzmann probability, $e^{-\beta\Delta E}$, where $\Delta E$ is the difference in energy between the new and old configurations (after and before the flip), and $\beta$ is the inverse of the temperature which measures the uncertainty or irrationality of individuals. We simulate both layers simultaneously in our two-layer model and calculate the energy differences for each layer. The energy difference for the leader layer is determined solely based on intra-layer triangles, while for the follower layer, it is calculated using both inter-layer and intra-layer triangles. As before, we accept the change of sign for randomly selected links in both layers if the energy difference for each layer either decreases or is accepted based on Boltzmann probability. Our simulation consists of $n^3$ Monte Carlo steps, and we use an ensemble size of $10^4$ to calculate standard deviations. The network size for our simulations is $n=50$; we also conducted simulations for larger network sizes, but the main results remained consistent.

    \begin{figure}[t]
        \centering
        \includegraphics[width=1\linewidth]{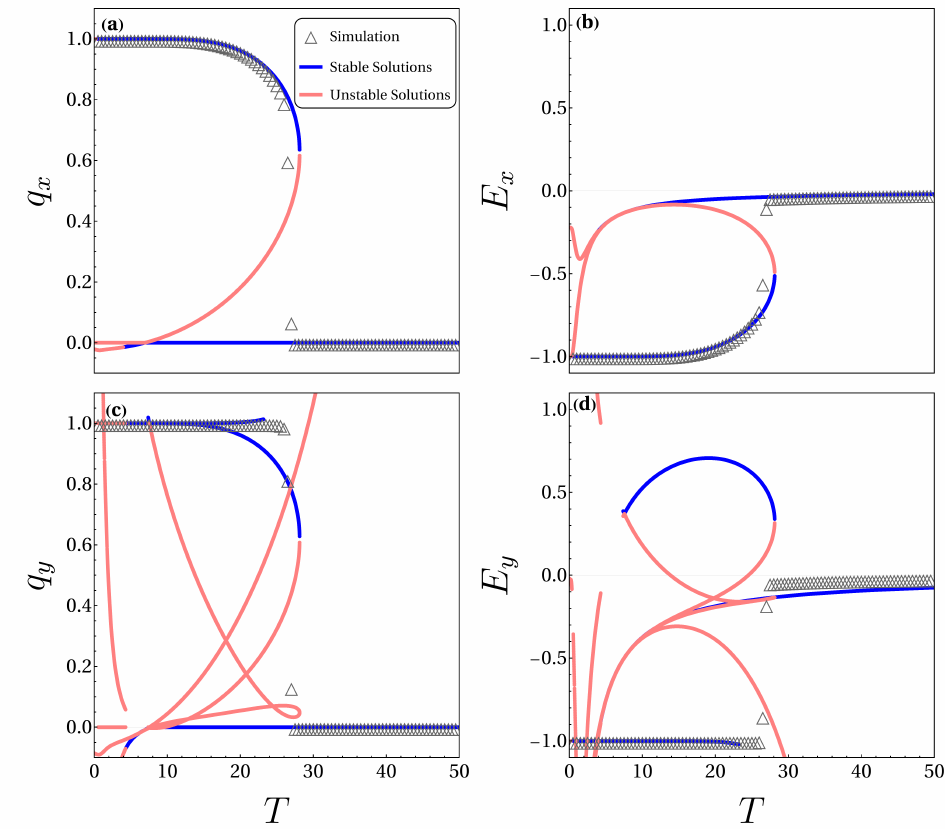}
        \caption{Comparison of our analytic mean-field approximation (blue and red solid lines) with simulations (triangles). The blue/red lines represent the stable/unstable solutions of our model self-consistence Eq.~\ref{selfcons}. The triangles represent the Monte Carlo simulation with the initial condition of all links having positive values. The first row plots the mean of two stars and energy for the leader layer, and the second row is related to the mean of two stars and the energy of the follower layer. The critical temperature is predicted with a mean-field approximation, which agrees with simulations. The number of nodes in all of our calculations is $n=50$.}
        \label{fig:main2}
    \end{figure}

    In Fig.~\ref{fig:main2}, we compare our Monte Carlo simulations with our analytical calculation for the mean of two-stars, as well as the energy of the leader layer ($x$) and the follower layer ($y$). The blue and red solid lines represent stable and unstable solutions, respectively. The stability of these solutions is discussed in the last paragraph of subsection \ref{first_sub_analysis}. The triangles in the figure represent results from the Monte Carlo simulations using an initial condition with all-positive links. Our analytical solution for the follower layer indicates a higher number of unstable solutions than the leader layer, leading to increased instability in the follower layer. The predicted critical temperature for the follower layer aligns well with our simulations.

    \begin{figure}[t]
         \centering
         \includegraphics[width=1\linewidth]{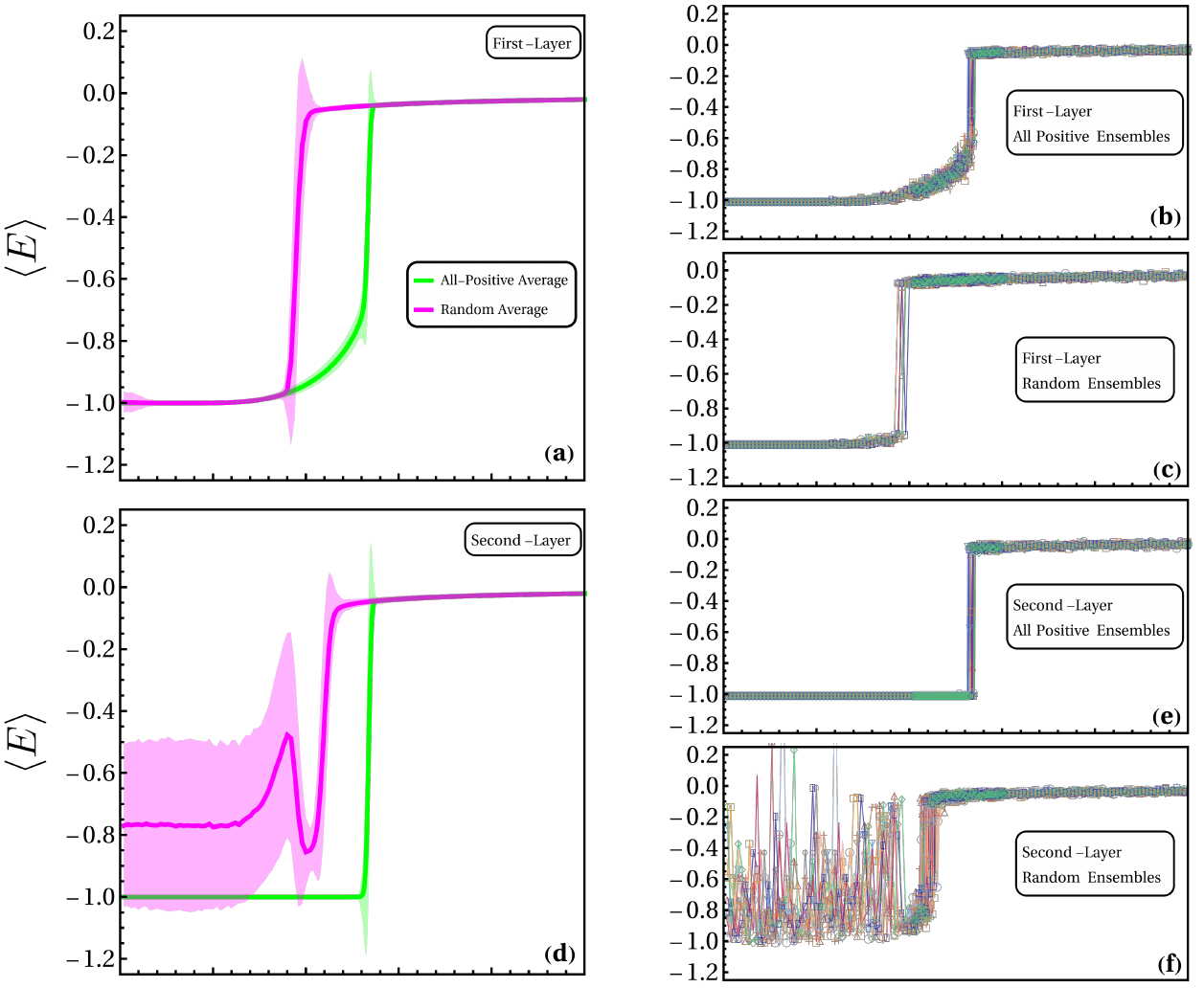}
         \caption{Comparison of Monte Carlo simulation with different boundary conditions for the leader and follower layers. All $10^4$ ensembles begin with heaven or random configurations. In panel (a), the average energy of heaven/random initial condition is shown with a solid green/magenta line, and the standard deviation is shown with a green/magenta shadow. Panels (b) and (c) depict ten ensembles' mean energy versus temperature for both initial conditions. Panel (d) exhibits the mean energy for the follower layer for different initial conditions. Ten ensembles' behaviors versus temperature for this layer are shown in panels (e) and (f). The number of nodes in all of our calculations is $n=50$.}
         \label{fig:main3}
     \end{figure}

    In Fig.~\ref{fig:main3}, panels (a) compare the mean energy $\langle E \rangle$ as a function of temperature, based on our Monte Carlo simulations with different initial conditions for the leader layer. The solid green and magenta lines represent the average energy of $10^4$ ensembles, while the light green and magenta shaded areas indicate the standard deviation for all-positive and random initial conditions. The initial condition indicates that each of our $10^4$ ensembles starts in heaven or a random state. The random initial state is generated with a probability of $1/2$ for each link, allowing it to be either positive or negative. We conduct the simulations in these panels so that each ensemble is independent of the others, starting from the same initial boundary condition—heaven or random. Both sets of initial conditions exhibited phase transitions at different critical temperatures. This behavior is associated with a well-known phenomenon in the literature of critical phenomena called the \enquote{hysteresis loop}, characteristic of first-order or discrete phase transitions \cite{sornette,tadic}. This phenomenon can be observed when the temperature trends increase (from zero to fifty, curved green arrow) or decrease (from fifty to zero, curved magenta arrow), independent of the initial boundary condition. However, to find the minimized energy states more efficiently through computation, it is more effective to begin with a decreasing trend using a random initial condition, followed by an increasing trend with all-positive initial conditions. On the left-hand side panels (b) and (c), we present ten ensemble behaviors from our simulations to enhance understanding. Panel (a) of Fig. \ref{fig:main3} closely resembles the work of Rabbani \etal \cite{fereshteh}.
    
    In panel (d) of Fig.~\ref{fig:main3}, we illustrate the follower layer's mean energy behavior versus temperature. In the all-positive initial condition (solid green line), the phase transition is sharper than the leader layer, and the critical temperature is identical. The most interesting aspect occurs with the random initial conditions for this layer (solid magenta). In this scenario, the follower layer exhibits non-monotonic behavior versus temperature and a small standard deviation around the critical temperature, indicated by the magenta shading. However, as the temperature decreases, the shadows expand and remain prominent as the temperature approaches zero. The high standard deviation indicates numerous local minima in the energy of the follower's layer, meaning that at non-zero temperatures, the system cannot escape from them. The phenomenon of the hysteresis loop can be seen in this layer for the increasing and decreasing trends in temperature (green and magenta curved arrows). In panels (e) and (f), we compared the behavior of ten ensembles without averaging procedures to better understand non-monotonic and large standard deviations.

    Figure \ref{fig:main3} (d) reveals an important observation: the heat capacity ($c_v = \partial \langle E \rangle / \partial T$) of the follower layer becomes negative below the critical temperature. The follower layer functions as an open system where its entropy can decrease. However, by the second law of thermodynamics, the overall entropy of the layers must still increase. This concept is similar to how a refrigerator operates. While a refrigerator reduces the entropy within its compartment by removing heat, it releases that heat into the environment, thus increasing the universe's entropy. This behavior demonstrates characteristics of open systems, which are similar to our model.

    In Fig.~\ref{fig:main4}, we examined the critical temperatures for both the leader and follower layers under conditions of an all-positive panel (a) and a random initial panel (b). For all-positive initial conditions, the critical temperatures for both layers are identical, panels (a). However, under random initial conditions, the critical temperature of the follower layer is higher than the leader layer, panels (b). This difference between critical temperatures for layers indicates that the follower layer collapses before the leader layer transitions, which can lead us to an interesting social interpretation.
	
    \section{Conclusion}

    \begin{figure}[t]
        \centering
        \includegraphics[width=1\linewidth]{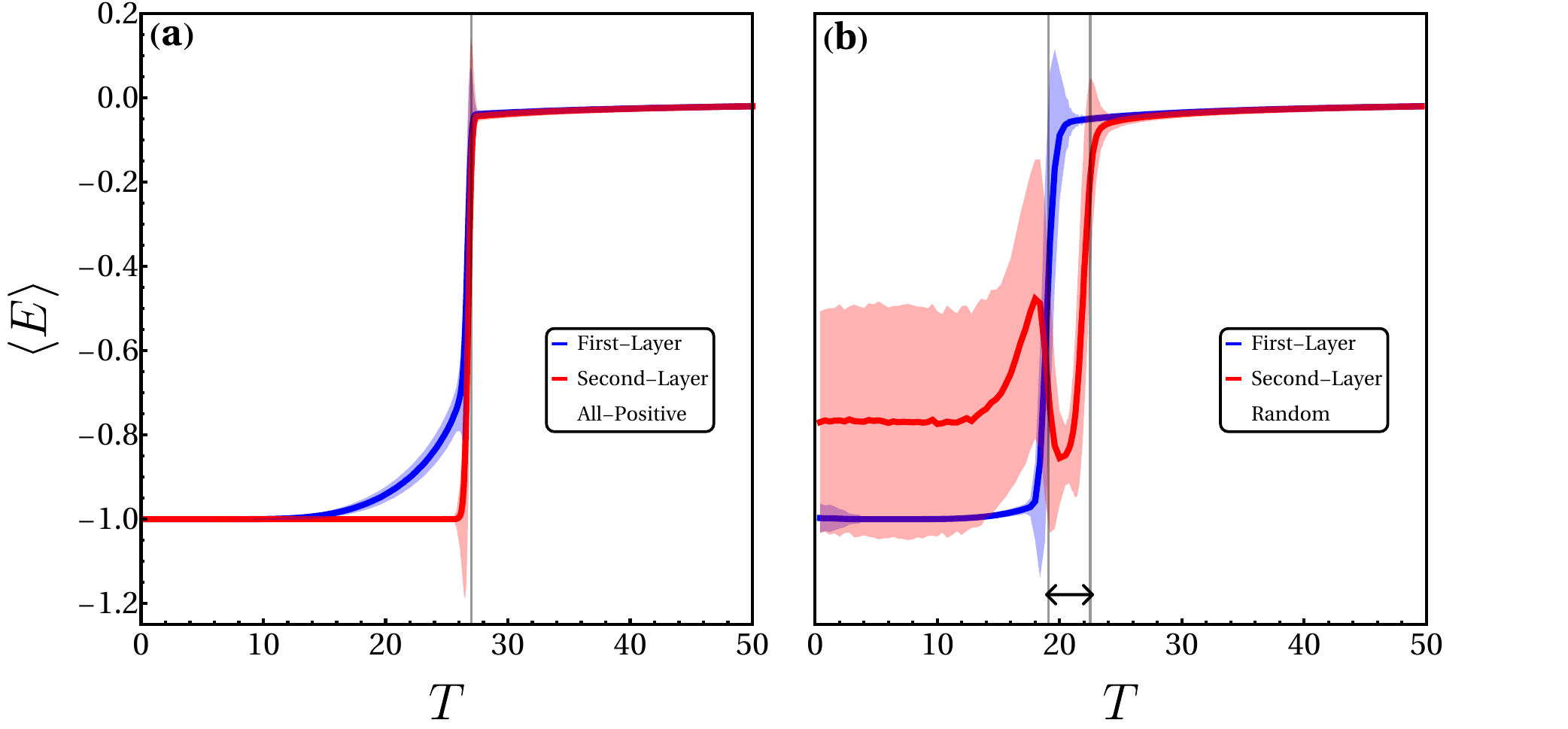}
        \caption{Comparison of the critical temperatures for both leader and follower layers under all-positive and random initial conditions is presented. In panel (a), which depicts the all-positive initial condition, the critical temperatures for both layers are identical. However, in panel (b), which represents random initial conditions, the critical temperature of the follower layer is higher than that of the leader layer. The simulations were conducted using $10^4$ ensembles, and the number of Monte Carlo steps is $n^3$. The number of nodes is set to $n=50$ in all our calculations.
}
        \label{fig:main4}
    \end{figure}

    The existence of hierarchy in societies of intelligent beings is undeniable \cite{mageeh,drews}. In human societies, some relationships are more important than others, such as relationships between close family and casual friends. As relationships within social networks evolve from simple social ties to more complex triplet dynamics, maintaining harmony among different types of interaction becomes increasingly challenging. In these situations, our attention often shifts to the leader-layer interactions (like family relationships), neglecting other relationships (follower-layer, like more casual friendships). The social models that encapsulate the concept of hierarchy by considering two or more types of social ties are more realistic. To our knowledge, there are extensions of balance theory models that have considered this concept \cite{gorski1,mohandas1,mohandas2}. These models have Heider balance interaction within layers and Ising-like interaction between layers. The inter-layer interactions are symmetric, meaning all agents are in identical social classes. Our model distinguishes followers and leaders, which means an asymmetric interaction between layers. Furthermore, all interactions within and between layers are Heider balance interactions. In other words, the balance of the follower layers depends on the balance of the leader layers, and the leader layer balance is independent of the follower layer. Begin with non-equilibrium rate equations and changing the approach to equilibrium shows that this asymmetrical dependence is evident in a large unstable solution observed in our analytical approach for the follower layer under critical temperature (Fig.~\ref{fig:main2}). Moreover, we verify these findings through numerous Monte Carlo simulations (Fig.~\ref{fig:main3}) and Fig.~\ref{fig:main4}). The instability of the follower layer suggests that reaching a balanced state may be fundamentally unattainable.

    \appendix
    \begin{appendices}
        \section{Deriving Self Consistence for Total and leader Layer}\label{appendix:self-con-fun}
    
        In this section, we derived the self-consistency equation for the leader layer, referred to as the leader layer (denoted as $x$), as well as for the total layers, denoted as $z$. We encountered terms involving arbitrary links, such as $x_{ij}$ or $z_{ij}$, which we encapsulated in $\mathcal{H}^x_{ij}$ and $\mathcal{H}^z_{ij}$. We separated the Hamiltonian terms that do not include these arbitrary links. By applying the mean-field approximation, which is summarized to $\langle AA\rangle\approx \langle A\rangle ^2$ for incidental quantity $A$, we formulated a self-consistency equation for the leader layer $x$ and the total layers $z$. So, all terms that have arbitrary links in the leader and total layers are
        \begin{equation}
            \mathcal{H}^x_{ij}=- x_{ij}{\sum_{k\neq{i,j}}x_{jk}x_{ki}},\quad
            \mathcal{H}^z_{ij}=- z_{ij}{\sum_{k\neq{i,j}}z_{jk}z_{ki}}.        
        \end{equation}
        The mean value $\langle x_{ij}\rangle$ and $\langle z_{ij}\rangle$ can be calculated as
        \begin{align}
            \langle{x_{ij}}\rangle &=\frac{1}{\mathcal{Z}_x}\sum_{\{x'\neq x_{ij}\}}e^{-\beta\mathcal{H}_x'}\sum_{x_{ij}={\left\{\pm{1}\right\}}}x_{ij}e^{-\beta\mathcal{H}^x_{ij}},\\
            \langle{z_{ij}}\rangle &=\frac{1}{\mathcal{Z}_z}\sum_{\{z'\neq z_{ij}\}}e^{-\beta\mathcal{H}_z'}\sum_{z_{ij}={\left\{0,\pm{1}\right\}}}z_{ij}e^{-\beta\mathcal{H}^z_{ij}},
        \end{align}
        In the equations above, we define $\mathcal{H}_z=\mathcal{H}'_z+\mathcal{H}^z_{ij}$ and $\mathcal{H}_x=\mathcal{H}'_x+\mathcal{H}^x_{ij}$. 
        Here, $\mathcal{H}'_z$ and $\mathcal{H}'_x$ represent terms that involve arbitrary links in the leader layer and the total layers, denoted as $x_{ij}$ and $z_{ij}$, respectively. The parameter $\beta$ represents the inverse of the temperature. It is important to note that the links in the leader layer take values from $x_{ij} \in \{-1, +1\}$, while the links in the total layer can take values from $z_{ij} \in \{-1, 0, +1\}$. By utilizing the definitions provided in Eq.~\ref{definitions}, we can calculate the mean of the links and two-stars for both the leader and total layers.
        
                \begin{align}
                    p_x &= \tanh(\beta (n-2) q_x),  \\
                    q_x &= \frac{\e^{\beta (p_x - 2 (n-3) q_x)} + \e^{\beta (2 (n-3) q_x + p_x)} - 2 \e^{-\beta p_x}}
                    {\e^{\beta (p_x - 2 (n-3) q_x)} + \e^{\beta (2 (n-3) q_x + p_x)} + 2 \e^{-\beta p_x}} \\
                    p_z &= \frac{\e^{\beta (n-2) q_z} - \e^{-\beta (n-2) q_z}}{\e^{-\beta (n-2) q_z} + \e^{\beta (n-2) q_z} + 1},  \\           
                \end{align}
                and the mean of two stars in the total layer is
                \begin{widetext}
                    \begin{equation}
                        q_z = 
                        \frac{
                        \e^{\beta (p_z - 2 (n-3) q_z)} + \e^{\beta (2 (n-3) q_z + p_z)} - 2 \e^{-\beta p_z}
                        }
                        {
                        \e^{\beta (p_z - 2 (n-3) q_z)} + \e^{\beta (2 (n-3) q_z + p_z)} 
                        + 2 \left( \e^{-\beta (n-3) q_z} + \e^{\beta (n-3) q_z} \right) 
                        + 2 \e^{-\beta p_z} + 1
                        }
                    \tag{A4}
                    \end{equation}
                \end{widetext}
                By putting the RHS of quantities of $p_x$ and $p_z$ in the function in the right-hand side of equations for $q_x$ and $q_z$, we derive the self-consistency equation of our model, which are
                \begin{equation}
                       \left\{
                       \begin{aligned}
                            q_{x} &=f(q_{x};\beta,n),\\
                            q_{z} &=g(q_{z};\beta,n).
                        \end{aligned}
                        \right.
                \end{equation}
                The function $f$ resembles the work of Rabbani et al. \cite{fereshteh}.
        
        The self-consistency equations mentioned above can be derived using a well-known method in quantum field theory. This approach involves adding a term equivalent to the mean quantity we want to examine to the Hamiltonian, along with the desired coupling constant. We then calculate the average of this quantity, the partition function, and the free energy. The mean quantity can be obtained by taking the derivative of the free energy with respect to the coupling constant. By considering the limit of the coupling constant as it approaches zero, we can observe the exact form of the mean quantity. We derive our model's self-consistency equations using this method. When we compare our results with the equation above, we find they are identical, confirming that our calculation is correct.

        \section{Mean-field approximation for energy of follower layer}\label{appendix:energy-follower-layer}
        The mean energy of triads for thermal balance theory calculated by Rabbani \ et \cite{fereshteh}. In this section, we want to perform the same calculation to find the mean energy of the follower layer from the leader and the total layers' energies, approximated with the mean-field method. We can write below the relation for total layer triads
        \begin{equation}\label{qx}
            \begin{aligned}
                -z_{ij}z_{jk}z_{ki} &= -\left(x_{ij}+y_{ij}\right) \left(x_{jk}+y_{jk}\right) \left(x_{ki}+y_{ki}\right)/8\\
                &= (-x_{ij} x_{jk} x_{ki} - x_{ij} y_{jk} y_{ki} - y_{ij} x_{jk} y_{ki}\\
                &- y_{ij} y_{jk} x_{ki} - y_{ij} x_{jk} x_{ki} - x_{ij} y_{jk} x_{ki}\\
                &- x_{ij} x_{jk} y_{ki} - y_{ij} y_{jk} y_{ki})/8.
            \end{aligned}
        \end{equation}
        The mean-field approximation of triplet interaction can be considered as below
         \begin{equation}
            \begin{aligned}
                 &\av{x_{ij} y_{jk} y_{ki}}=\av{x_{ij}}\av{y_{jk} y_{ki}} = p_x q_y,\\
                 &\av{y_{ij} x_{jk} x_{ki}}=\av{y_{ij}}\av{x_{jk} x_{ki}} = p_y q_x\\
                 &\av{y_{ij} x_{jk} y_{ki}}=\av{ y_{ij} y_{jk} x_{ki}}=\av{y_{ij}}\av{x_{jk} y_{ki}} = p_y q_{xy},\\
                 &\av{x_{ij} y_{jk} x_{ki}}=\av{ x_{ij} x_{jk} y_{ki}}=\av{x_{ij}}\av{y_{jk} x_{ki}} = p_x q_{xy},\\    
            \end{aligned}
        \end{equation}
        By knowing that $ q_{xy} \approx p_x p_y $ and defining $ E_z \equiv -\langle z_{ij} z_{jk} z_{ki} \rangle $,  $E_x \equiv -\langle x_{ij} x_{jk} x_{ki} \rangle$, we can express the energy of the total system as
        \begin{equation}
            \begin{aligned}
                 E_y &\equiv -\av{y_{ij} y_{jk} y_{ki}} \\
                 &=  \big[(8E_z - E_x) + p_x( q_y + 2 p_x p_y) + p_y( q_x + 2 p_x p_y)\big]
            \end{aligned}
        \end{equation}
	\end{appendices}

        \bibliographystyle{apsrev4-1}
        \bibliography{reference-TLBT}       

\end{document}